# Destruction of graphene by metal adatoms

## D. W. Boukhvalov and M. I. Katsnelson

*Institute for Molecules and Materials, Radboud University Nijmegen, Heyendaalseweg 135, 6525 AJ Nijmegen, The Netherlands*

The formation energies for mono- and bivacancies in graphene in the presence of adatoms of various metals and small metallic clusters have been calculated. It is shown that transition metal impurities, such as iron, nickel and, especially, cobalt reduce dramatically the vacancy formation energies whereas gold impurities have almost no effect on characteristics of the vacancies. This results highlight that special measures are required in order to protect graphene from damage by transition metal leads.

Graphene [1,2,3] is now considered as a prospective material for ``beyond-silicon'' electronics. In particular, its potential applications in various spintronics devices have been studied both experimentally [4] and theoretically [5,6]. In particular Cr/Au gates are used in these fare mentioned experimental prototypes of graphene-ferroelectric memory and other graphene-based devices [7,8]. The physics of the processes involved at the interface between graphene and the metallic electrodes remains unclear. Unexpectedly strong spin-flip relaxation [4] and even polarization of π-orbitals of carbons in graphene on cobalt [9] have been observed. Electronic structure of graphene with metal leads or on metal substrates [6,10,11,12] needs further investigation. In addition even the ability to obtain clean graphite, free from impurities of 3*d* elements, especially, iron [13] is a significant problem. After the preparation of graphene from graphite these impurities can effect essential properties of the carbon layer. Metal adatoms can induce major  structural reconstruction of graphene, as one can assume from experiments on cut-off graphene by metallic nanoparticles [14,15]  and atomic force microscope (AFM) tip [16,17]. These methods can be successfully used to derive graphene nanoribbons by the unzipping of carbon nanotubes [18,19]. The destruction of fullerenes with formation of carbide phases [20,21] as well as their formation at the border of nickel electrodes and carbon-contained nanostructures [22] demonstrate relevance of metal-graphene interactions and resulting problems.

The energetics and geometry of individual metal adatoms on graphene and the modification of defects in graphene by atamoms of gold already has been studied [23,24,25]. Previously we have shown that different types of defects in graphene are centers of chemical activity [26]. It was found also[27] reported that adatoms of iron can play an important role in the formation of vacancies in fullerenes $C_{60}$. Here we investigate how presence of metallic impurities in forms of single atoms, pairs of atoms and atomic tetrahedra can dramatically change the formation energy of mono- and bivacancy in graphene.

Our calculations have been carried out using the SIESTA code [28,29] with the generalized gradient approximation (GGA) [30] to DFT and Troullier-Martins [31] pseudopotentials. We used energy mesh cutoff of 400 Ry, and 12×12×2 *k*-point mesh in the Monkhorst-Park scheme [32]. During the optimization, the electronic ground states were self-consistent by using the norm-conserving pseudopotentials for the cores and a double-ζ plus polarization basis of localized orbitals for carbon and the metals. Optimization of the bond lengths and total energies was performed with an accuracy 0.04 eV / Å and 1 meV, respectively. This technical parameters of the computations are the same as those report in our earlier studies on carbon-iron systems[27,33]. The vacancy formation energies were calculated as $E_{vacancy} = E_{graphene\ with\ vacancy\ +\ metal\ adatoms} - E_{graphene\ +\ metal\ adatoms} - E_{carbon}$ where $E_{carbon}$ is the total energy per single carbon atom in pristine graphene.

We will consider five different species of adatoms: aluminium, which can diffuse from the dielectric substrate $Al_2O_3$ [4]; iron, which is the most common metallic impurity found in graphite [13]; cobalt, which was used for electrodes in Refs. 4,9; nickel, which was used as a substrate for the growth of graphene [34,38]; and gold, also used for contacts in many experiments with graphene. We will consider three types of metallic impurities: single adatom (see Fig. 1a), pair of adatoms situated parallel to the graphene plane (Fig.2a) and tetrahedron oriented as shown in Fig. 2b. The latter situation models an interaction of graphene with microscope tip or with structural inhomogeneities on metallic electrodes. It is worthwhile to note that, as follows from calculations, inverse orientation of the tetrahedron (face to the graphene plane) will have total energies roughly 1 eV higher.

For the cases of single metallic adatoms we have obtained cohesive energies and equilibrium metal-carbon distances very close to those by Chan et al.[23]. Formation energies for mono- and bivacancies are shown in Fig. 3. For the cases of adatoms of gold, the vacancy formation energy, 5 to 6 eV, is close to that in graphene without impurities [35]. For transition metals, the vacancy formation energies are much lower, from 1 to 3 eV, which meance the vacancy formation in the presence of these metals highly probable in particular if graphene is warmed during the experiments. In contrast gold is very weakly bonded with graphene (cohesive energy is about 0.01 eV, in agreement with the previous work [23]) and the corresponding out-of-plane distortions of graphene are almost negligible (about 0.16 Å). In the presence of vacancy, the main effect of the metals is just filling the void by gold atom which slightly decreases the vacancy formation energy. For adatoms of transition metals, the bonding is much stronger (the cohesive energy is 0.2 to 0.3 eV) and the charge transfer is essential. As well as for pure graphene [35], doping diminishes the vacancy formation energy. Whereas the bivacancy formation energies are almost the same in the presence of Fe, Co, and Ni, the monovacancy formation energy is the lowest in the case of Co.

During the calculations we have also taken into account spin polarization. Nickel turns out to be non-magnetic, both on pristine graphene and for mono- and bivacancies. For iron and cobalt the magnetic moments are parallel and antiparallel to the moment of monovacancy, respectively (bivacancy is always non-magnetic, due to the absence of dangling bonds). Interestingly, the electronic structures of the transition metals on a monovacancy (Fig. 4) are rather similar to that those of the corresponding carbides [37].

It can be seen in Fig. 3 that the same tendencies are correct for the cases of adatom dimers (Fig.

2a) and tetrahedra (Fig. 2 b,c). In all cases, the monovacancy formation energy is maximal for gold and minimal for cobalt. For bivacancy geometric factors becomes more important, especially, for the case of tetrahedron. For iron, cobalt and nickel, the distortions of the tetrahedron into carbide-like configuration Me$_3$C are very strong (Fig. 2c) whereas in the case of no carbides are formed.

It follows from our calculations that a special protection of graphene from the destructive effects of transition metals is desirable. For example, one can cover electrodes by monolayer of gold. To model this we have calculated vacancy formation energies in the presence of tetrahedron of Fe, Co, Ni, with a replacement by the metal atom nearest to graphene by the atom of gold. The monovacancy formation energies in these cases are 5.11 eV, 4.84 eV, and 5.23 eV, respectively, which are very close to the values for a pure gold tetrahedron, 5.78 eV. A strong spin polarization of graphene on cobalt has been observed whereas no polarization has been observed in graphene [9,38] on nickel covered by a monolayer of gold. On the other hand, our results show that cobalt strongly destabilize graphene lattice and thus could be an appropriate metal for the slicing of nanoribbons and nanoflakes with a given shape from graphene which, may be interesting in themselves [39].

**Acknowledgment** The work is financially supported by Stichting voor Fundamenteel Onderzoek der Materie (FOM), the Netherlands. Authors acknowledge Prof. A. E. Rowan for them careful reading of the manuscript.

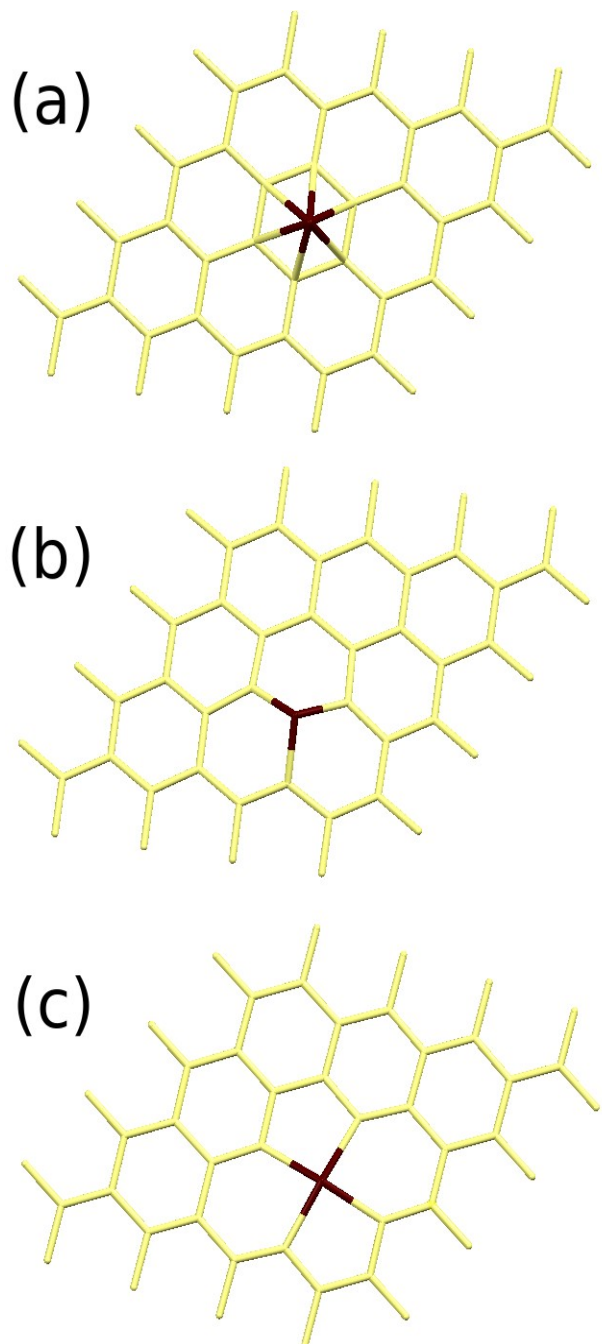

**Figure1.** (color online) Optimized structures for the cases of single iron adatom on pristine graphene (a), graphene with monovacancy (b), and bivacancy (c).

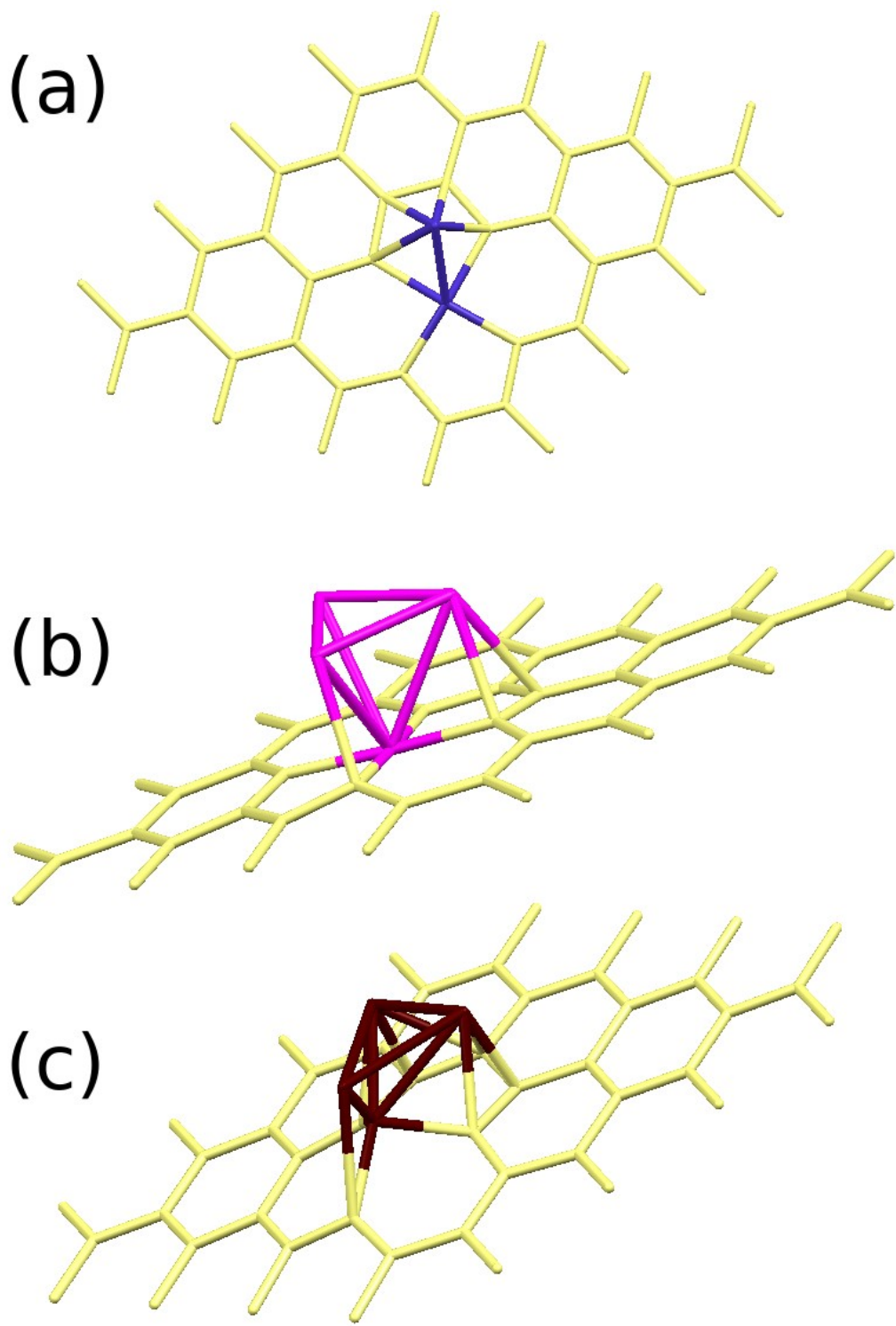

**Figure 2** (color online) Optimized structures for the cases of cobalt dimer (a), gold (b) and iron (c) tetrahedra on graphene with bivacancy.

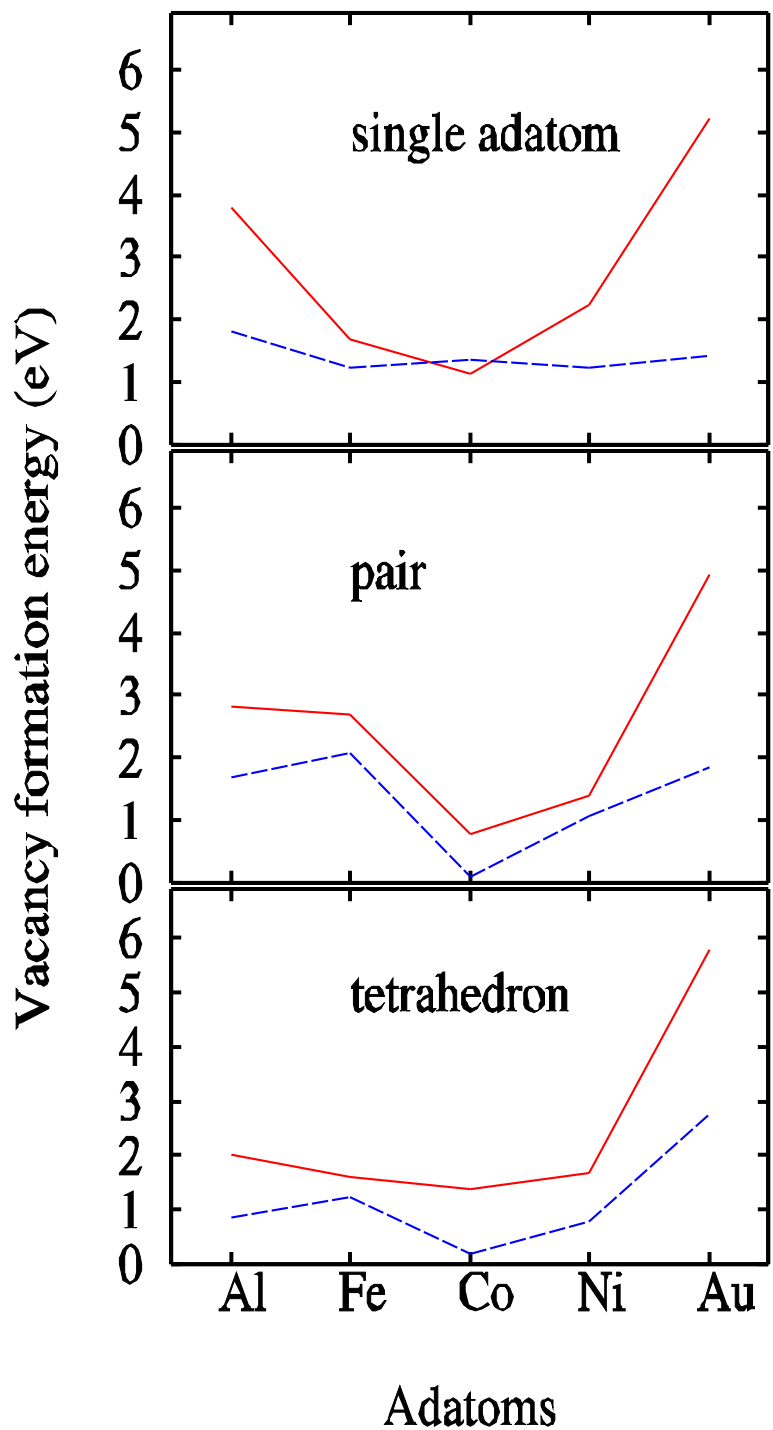

**Figure 3** (color online) Formation energies per removed carbon atom for monovacancy (solid red line) and bivacancy (dashed blue line) in the presence of different adatoms on graphene.

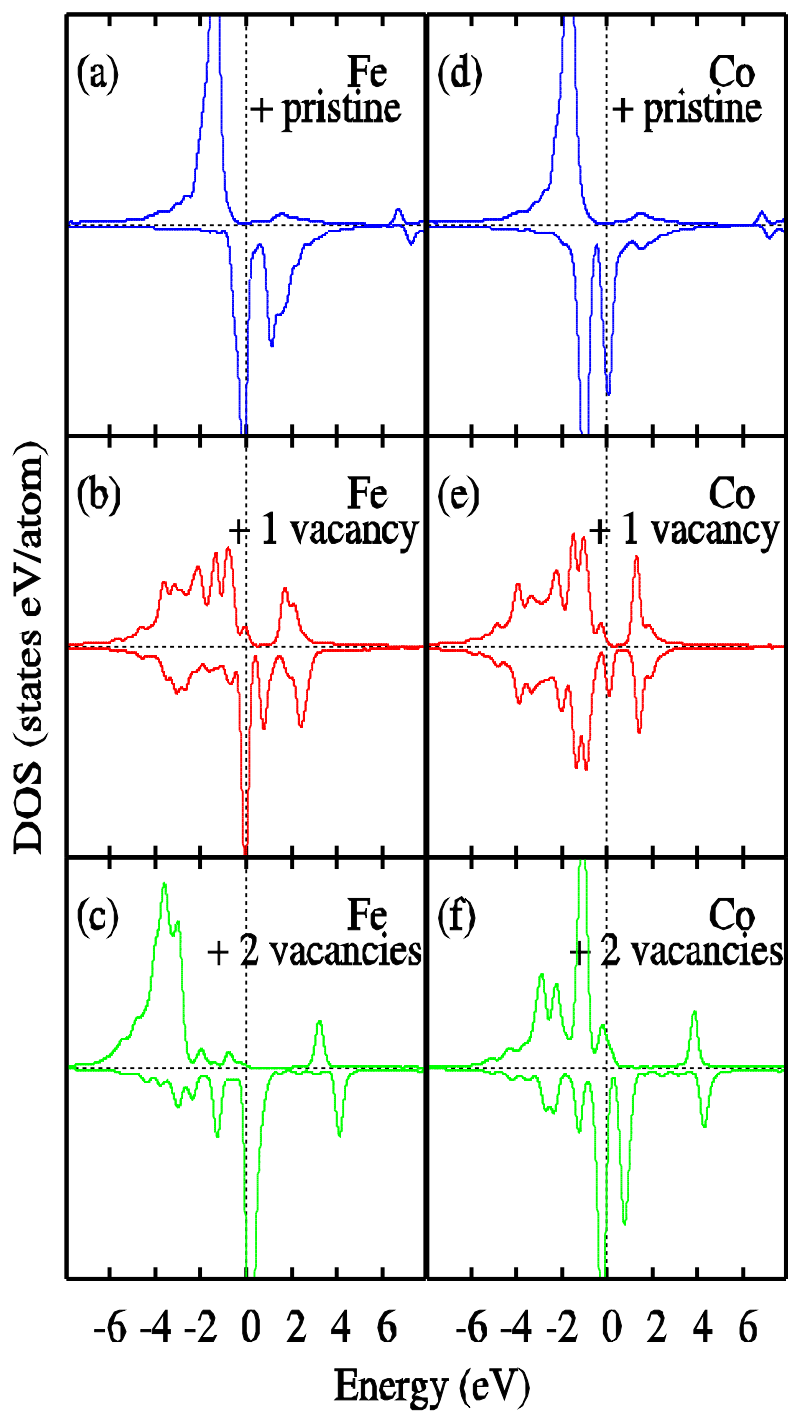

**Figure 4** (color online) Spin polarized densities of states (up - majority, down - minority spins) for $3d$ orbitals of single adatoms on pristine graphene and graphene with vacancies (see Fig. 1).